\documentclass[aps,floatfix,nofootinbib,showpacs,twocolumn]{revtex4}

\usepackage{graphicx}
\usepackage{dcolumn}
\usepackage{amsmath}

\usepackage{color}
\definecolor{purple}{rgb}{0.5,0,0.5}
\definecolor{blue}{rgb}{0.0,0,0.9}

\begin{document}


\title{%
%
Gauge invariance of a critical number of flavours in QED3}

\author{A.~Bashir}
\affiliation{Instituto de F\'{\i}sica y Matem\'aticas, Universidad Michoacana de San Nicol\'as de Hidalgo, Apartado Postal 2-82, Morelia, Michoac\'an 58040, M\'exico}

\author{A.~Raya}
\affiliation{Instituto de F\'{\i}sica y Matem\'aticas, Universidad Michoacana de San Nicol\'as de Hidalgo, Apartado Postal 2-82, Morelia, Michoac\'an 58040, M\'exico}

\author{S.~S\'anchez-Madrigal}
\affiliation{Instituto de F\'{\i}sica y Matem\'aticas, Universidad Michoacana de San Nicol\'as 
de Hidalgo, Apartado Postal 2-82, Morelia, Michoac\'an 58040, M\'exico}

\author{C.\,D.~Roberts}
\affiliation{Physics Division, Argonne National Laboratory, Argonne,
Illinois 60439, USA}
\affiliation{Department of Physics, Peking University, Beijing 100871, China}

\begin{abstract}
The fermion propagator in an arbitrary covariant gauge can be obtained from the Landau gauge result via a Landau-Khalatnikov-Fradkin transformation.  This transformation can be written in a practically useful form in both configuration and momentum space.  It is therefore possible to anticipate effects of a gauge transformation on the propagator's analytic properties.  These facts enable one to establish that if a critical number of flavours for chiral symmetry restoration and deconfinement exists in noncompact QED3, then its value is independent of the gauge parameter.  This is explicated using simple forms for the fermion-photon vertex and the photon vacuum polarisation.  The illustration highlights pitfalls that must be avoided in order to arrive at valid conclusions.  Landau gauge is seen to be the covariant gauge in which the propagator avoids modification by a non-dynamical gauge-dependent exponential factor, whose presence can obscure truly observable features of the theory. 
\end{abstract}


\maketitle

\section{Introduction}
\label{sec:one}
Quantum electrodynamics in three dimensions (2 spatial and 1 temporal - QED3) is of perennial interest to those concerned with confinement and dynamical symmetry breaking in quantum field theory because the quenched theory possesses a nonzero string tension \cite{Gopfert:1981er} and this feature persists in the unquenched theory if massive fermions circulate in the photon vacuum polarisation \cite{Burden:1991uh}.  That mass can either be explicit or dynamical in origin.  Since the theory is super-renormalisable, it has a well-defined chiral limit and therefore admits the possibility and study of dynamical mass generation.  This increases its quantitative similarity with QCD because dynamical chiral symmetry breaking (DCSB) explains \cite{Roberts:2007ji} the origin of constituent-quark masses  and underlies the success of chiral effective field theory.  In parallel with its relevance as a tool through which to develop insight into aspects of QCD, QED3 is also of interest in condensed matter physics as an effective field theory for high-temperature superconductors \cite{Franz:2002qy,Herbut:2002yq,Thomas:2006bj} and graphene \cite{Novoselov:2005kj,Gusynin:2007ix}.  

A perspective on confinement is laid out in Refs.\,\cite{Krein:1990sf,Roberts:1994dr,Roberts:2000aa,Roberts:2007ji}, which explain that a sufficient condition for the confinement of a given elementary excitation is the absence of a K\"all\'en-Lehmann representation for the associated $2$-point Schwinger function.  In the noncompact formulation of QED3 there are indications that fermion confinement and DCSB go hand-in-hand \cite{Maris:1995ns,Bashir:2008fk}.\footnote{Features of confinement in the gauge sector of a compact formulation of QED3 are discussed, e.g., in Ref.\,\protect\cite{Chernodub:2002gp}.}  

A fascination with this stems from an anticipation that QED3 possesses a critical number of flavours, $N_f^c$, above which DCSB, and therefore confinement, is impossible \cite{Appelquist:1988sr}.  This feature makes the theory interesting as a tool with which to explore technicolour extensions of the Standard Model \cite{Sannino:2008kg} and possibly relevant to the phase diagram of QCD in the plane formed by the coupling and the number of quark flavours; e.g., Ref.\,\cite{Gies:2005as,Kurachi:2006mu}.  

The study of lattice-regularised QED3 suggests $N_f^c>1$ \cite{Hands:2004bh}, with a recent simulation hinting at $N_f^c \sim 1.5$ \cite{Strouthos:2008kc}.  The latter is curious because it was argued in Ref.\,\cite{Appelquist:1999hr} that $N_f^c \leq 3/2$, although this constraint has been challenged \cite{Mavromatos:2003ss}.  However, in connection with lattice simulations it should be noted that an impediment to reliable results is the mass hierarchy feature of QED3; viz., any dynamically generated mass-scale is at least one order of magnitude smaller than the natural scale, which is set by the dimensioned coupling $e^2$ (see, e.g., Ref.\,\cite{Bashir:2005wt}).\footnote{This is not the case for QCD, which possesses a dimensionless running coupling that evolves with respect to a mass-scale whose magnitude is characteristic of all dynamically generated mass-dimensioned quantities in the theory.}  
As one would readily anticipate, this accentuates the impact of finite volume artefacts and can lead to underestimation of $N_f^c$ because the signal for DCSB is lost once the magnitude of this effect falls below $1/L$, where $L$ is the lattice length-scale \cite{Goecke:2008zh}.  

It was recently demonstrated \cite{Bashir:2008fk} that noncompact QED3 can possess a critical number of flavours if, and only if, the fermion wave function renormalisation and photon vacuum polarisation are homogeneous functions at infrared momenta when the fermion mass function vanishes.  The Ward identity entails that the fermion-photon vertex possesses the same property and ensures a simple relationship between the homogeneity degrees of each of these functions.  These exact results were illustrated using a simple model for the photon vacuum polarisation and one for the fermion-photon vertex.  The existence and value of $N_f^c$ are naturally contingent upon the precise form of the vertex.  

It was also argued that should a critical number of flavours exist, then any discussion of its gauge dependence is moot because Landau gauge occupies a special place in gauge theories.  It is the gauge in which any sound \textit{Ansatz} for the fermion-photon vertex can most reasonably be described as providing a pointwise accurate approximation.  The vertex in any other gauge should then be defined as the Landau-Khalatnikov-Fradkin (LKF) transform \cite{LK56,Fr56,JZ59,BZ60} of the Landau gauge \emph{Ansatz}.  The sensible implementation of this procedure guarantees gauge covariance and hence obviates any question about the gauge dependence of gauge invariant quantities.  However, as we shall see, a deft hand is necessary when employing the LKF transformations.

We return to the study of QED3 in order to demonstrate gauge invariance in connection with the critical number of flavours for deconfinement and chiral symmetry restoration.  Notably, these transitions are coincident in Landau gauge \cite{Bashir:2008fk} and that should not change following a gauge transformation.  We recapitulate on those aspects of QED3 necessary for our discussion in Sec.\,\ref{QED3recap}; describe and explain our findings in Sec.\,\ref{sec:results}; and provide a summary and perspective in Sec.\,\ref{sec:close}.

\section{QED3 and the LKF transformations}
\label{QED3recap}
In three dimensions it is possible to work with two-component spinors.  However, in that formulation a mass term of any origin is parity-odd.  This problem may be avoided by employing four-component spinors and a $4\times 4$ representation of the Clifford algebra: $\{\gamma_\mu,\gamma_\nu\} = 2\delta_{\mu\nu}$; e.g., the set $\{\gamma_1,\gamma_2,\gamma_4\}$, taken from four dimensional theories, with $\gamma_5:=-\gamma_1\gamma_2\gamma_4$.  There are then two different mass terms; viz., ``$m\bar\psi\psi$'' and ``$m\bar \psi \frac{1}{2}[\gamma_3,\gamma_5]\psi$.''  The former is analogous to the natural form in four dimensions and it is invariant under parity transformations.  Hence we use it to define the theory.

The study of confinement and DCSB can be pursued through the gap equations for the fermion and photon; namely, in a theory with $N_f$ fermions of mass $m$,
\begin{eqnarray}
S(p)^{-1} &=&  i\gamma\cdot p\, A(p)+B(p)\,,\\
&=& [i\gamma\cdot p + M(p)]/Z(p)\,, \label{MassFn}\\
&=& i\gamma\cdot p + m+\Sigma(p)\,,\\
\Sigma(p) &  = & \displaystyle e^2 \int\frac{d^3q}{(2\pi)^3} D_{\mu\nu}(p-q) \gamma_\mu S(q) \Gamma_\nu(q,p) \,,
\label{gendse}
\end{eqnarray}
where, with $q_\pm=q\pm k/2$,
\begin{eqnarray}
D^{-1}_{\mu\nu}(k) &=&\left[\delta_{\mu\nu} - (1-1/\xi)k_\mu k_\nu\right] +\Pi_{\mu\nu}(k),\label{Dprop}\\
\Pi_{\mu\nu}(k) &= &\left[k^2\delta_{\mu\nu} - k_\mu k_\nu\right] \Pi(k) =: T_{\mu\nu}(k)\, k^2\, \Pi(k), \label{Pimn}\\
\nonumber & = & - N_f \, e^2  \int\frac{d^3q}{(2\pi)^3} {\rm tr}\, \gamma_\mu S(q_+) \Gamma_\nu(q_+,q_-) S(q_-). \\ \label{PiDSE}
\end{eqnarray}
We observe that Landau gauge is obtained with $\xi=0$ in the photon propagator; massless QED3 is straightforwardly defined by setting the Lagrangian mass $m=0$; and the quenched theory is obtained by writing $\Pi(k)\equiv 0$.  Furthermore, as mentioned above, $e^2$ in QED3 has unit mass-dimension and, without loss of generality \cite{Bashir:2008fk}, one may work with $e^2=1$.  All mass-dimensioned quantities are then measured in terms of $e^2$. 

In principle one can evaluate the chiral-limit fermion propagator to any finite order in perturbation theory.  At O$(\alpha^2)$, $\alpha=e^2/(4\pi)$, \cite{Bashir:2000ur,Bashir:2004hh}:
\begin{equation}
\label{Apert}
%
A(p) = 1 + \xi \frac{\pi \alpha}{4 p} + \xi^2 \frac{\alpha^2}{4 p^2} \left[\frac{\pi^2}{4}-1\right] - \frac{3 \alpha^2}{4 p^2} \left[\frac{\pi^2}{4}-\frac{7}{3}\right],
\end{equation}
from which it is clear that there is no covariant gauge in which $A(p)\equiv 1$ is the fully nonperturbative solution.  On the other hand, the choice of Landau gauge suppresses the leading order contribution completely.\footnote{In Landau gauge the one-loop contribution to $A(p)$ vanishes in any number of dimensions in any renormalisable gauge theory \protect\cite{Pascual:1984zb,Davydychev:2000rt}.}  Hence, Eq.\,(\ref{Apert}) is most accurate for $\xi=0$, in which case we anticipate that it is reliable for 
\begin{equation}
\label{Aaccurate}
\frac{p}{e^2} \gg  \frac{\surd 3}{8 \pi} \left[\frac{\pi^2}{4}-\frac{7}{3} \right]^\frac{1}{2} \approx \frac{1}{40}\,.
\end{equation}
The accuracy deteriorates linearly with $\xi$: in Feynman gauge, $\xi=1$, the series can only be reliable for $p/e^2 \gg 1/4$.  That these estimates are reliable is apparent, e.g., from Ref.\,\cite{Bashir:2005wt}.

From the configuration-space Landau-gauge fermion propagator one obtains the form in another covariant gauge through the LKF transformation
\begin{equation}
\label{LKFx}
S(z;\xi) = S(z;\xi=0)\, {\rm e}^{- \varsigma |z|},\; \varsigma= \xi \frac{e^2}{8 \pi}= \frac{1}{2} \xi \alpha\, .
\end{equation}
This expression, in combination with Eqs.\,(\ref{Apert}), (\ref{Aaccurate}) and the unquenched result \cite{Burden:1991uh,Bashir:2008fk}: \mbox{$0\leq A(p=0;\xi=0)<\infty$}, highlights a particular feature of Landau gauge: it is that covariant gauge in which the infrared behaviour of the fermion propagator is neither enhanced nor suppressed by a non-dynamical gauge-dependent exponential factor.  

Owing to the mass-dimension of the coupling, the gauge-dependent transformation parameter, $\varsigma$ in Eq.\,(\ref{LKFx}), also has unit mass dimension.  Its scale is set by that of $\alpha$, which is of the same magnitude or larger than any dynamical mass scale the theory might generate. Confinement and DCSB are infrared effects; viz., they are apparent at a momentum-scale less than that which characterises a given theory.  It is thus clear at the outset that $\varsigma$ can potentially mask these effects in any gauge other than $\xi=0$.  However, as we shall see, it cannot destroy them: they will always be exposed in a careful analysis.

For $\xi>0$ one can convert Eq.\,(\ref{LKFx}) into a practical transformation of the fermion propagator in momentum space; i.e., from a Landau gauge solution 
\begin{equation}
\label{Ssigma}
S(p) = -i \gamma\cdot p \, \sigma_V(p^2;\xi=0) + \sigma_S(p^2;\xi=0)
\end{equation} 
one may obtain the form in a different covariant gauge through the operations \cite{Bashir:2005wt,Bashir:2000ur,Bashir:2002sp,Bashir:2004yt}
\begin{eqnarray} 
\nonumber
\sigma_V(p;\xi) & =& \frac{\varsigma}{\pi p^2}\int_0^\infty \!\! dk\, k^2 \sigma_V(k;0) \\
&& \times  
\left[\frac{1}{\lambda^ -}+\frac{1}{\lambda^+}+\frac{1}{2kp} 
\ln{  {\frac{\lambda^-}{\lambda^+}}  } 
\right], \label{gistV}\\
\sigma_S(p;\xi) &=&  
\frac{\varsigma}{\pi p}\int_0^\infty \!\! dk\, k\ \sigma_S(k,0) 
\left[\frac{1}{\lambda^-}-\frac{1}{\lambda^+}\right]\!, \label{gist} 
\end{eqnarray}
where $\lambda^\pm = \varsigma^2 + (k\pm p)^2$.  Focusing on the right-hand-sides of Eqs.\,(\ref{gistV}) and (\ref{gist}), it will readily be apparent that the limit $\varsigma\to 0^+$ (i.e., $\xi\to 0^+$) is well defined and yields $\sigma_V(p,0)$ and $\sigma_S(p,0)$.  This fact establishes self-consistency of the momentum-space integral transform.

As illustrated by Eq.\,(\ref{Apert}), on the domain of momenta for which a perturbative calculation of the fermion propagator is accurate, which is the ultraviolet domain in QED3, the passage between $\xi>0$ and $\xi < 0$ is straightforward because the perturbative expansion is analytic at $\xi=0$.  This simplicity is also apparent in the $z\simeq 0$ behaviour of Eq.\,(\ref{LKFx}).  

However, for $\xi = - |\xi| < 0$ it is less straightforward to obtain the dressed momentum space propagator from the nonperturbative Landau gauge result.  One cannot simply employ Eqs.\,(\ref{gistV}), (\ref{gist}).  This is evident because they are both odd under $\xi \to -\xi$, a property not shared by the perturbative expansion in Eq.\,(\ref{Apert}).  The breakdown originates in a failure for $\xi<0$ of the conditions used in deriving Eqs.\,(\ref{gistV}), (\ref{gist}).  Nonetheless, an inverted procedure is available; namely, $\sigma(p;-|\xi|)$ are those functions for which
\begin{eqnarray} 
\nonumber
\sigma_V(p;0) & =& \frac{|\varsigma|}{\pi p^2}\int_0^\infty \!\! dk\, k^2 \sigma_V(k;-|\xi|) \\
&& \times  
\left[\frac{1}{\lambda^ -}+\frac{1}{\lambda^+}+\frac{1}{2kp} 
\ln{  {\frac{\lambda^-}{\lambda^+}} } 
\right], \label{gistVm}\\
\sigma_S(p;0) &=&  
\frac{|\varsigma|}{\pi p}\int_0^\infty\!\! dk\, k\, \sigma_S(k,-|\xi|) 
\left[\frac{1}{\lambda^-}-\frac{1}{\lambda^+}\right]\!. \label{gistm} 
\end{eqnarray}
If these integral equations should prove impractical when applied to the solution obtained in a given truncation, then one may return to configuration space and employ Eq.\,(\ref{LKFx}).  Importantly, Eq.\,(\ref{LKFx}) shows that the gauge dependence of the propagator is analytic at all length scales.  However, in the infrared; i.e., for $|z|>1/\varsigma$, a large-order power series in $\xi$ is required to accurately represent that gauge parameter dependence.  It is on this domain, which corresponds to $p<\varsigma \approx \xi/25$, that Eqs.\,(\ref{gistVm}) and (\ref{gistm}) might provide an awkward tool.


In closing this section we judge it important to recall that the momentum-space location of a physical mass pole in the fermion propagator is gauge independent.  This owes to a cancellation between the gauge-dependent parts of the fermion's vector and scalar self-energies which can only occur at the pole position; see., e.g., Refs.\,\cite{Aitchison:1983ns,Elias:1985qp,Johnston:1986tb}.  Conversely, the \emph{absence} of a gauge-independent mass pole is not something that the LKF transformation can change \cite{Krein:1990sf}.

\section{Confinement and DCSB}
\label{sec:results}
These phenomena were studied in Ref.\,\cite{Bashir:2008fk} via two order parameters.  That for DCSB was simply
\begin{eqnarray}
\label{rhoDCSB}
\rho(N_f)= \frac{M(p=0;N_f)}{M(p=0;N_f=1)}\,,
\end{eqnarray}
where the mass function is defined in Eq.\,(\ref{MassFn}).  Naturally, one could also use an equivalent order parameter; namely, the vacuum fermion condensate:
\begin{equation}
\label{qbarq}
-\langle \bar \psi \psi \rangle_{N_f} = {\rm tr}_{\rm D}\int\frac{d^3q}{(2\pi)^3} \, S(q)\,,
\end{equation}
evaluated in the chiral limit, and we also consider that herein.  

While these order parameters for DCSB are familiar, that for confinement might require a short explanation.   Working with Eq.\,(\ref{Ssigma}), then in the absence of confinement ($x=p^2$)
\begin{equation}
\frac{d^2}{dx^2} \, \sigma_V(x) > 0 \,, \; \forall x>0\,.
\end{equation}
On the other hand, $S(p)$ describes a confined excitation
\begin{equation}
\label{RPbroken}
\mbox{if}\;\exists x_c>0:\; \left.\frac{d^2}{dx^2} \, \sigma_V(x)\right|_{x=x_c} \! =0.
\end{equation}
The location of a minimum in $\sigma_V^\prime(x)$, $x_c$, can therefore be used to define an order parameter for deconfinement; viz.,\footnote{The location of the minimum positioned farthest from $x=p^2=0$ should be used.  In an asymptotically free theory, if there are any minima at all, then there will be one most distant.}
\begin{equation}
\label{Kconf}
\kappa(N_f) = \frac{x_c(N_f)}{x_c(N_f=1)}.
\end{equation}
These statements are associated with the realisation of confinement through a violation of the axiom of reflection positivity.  Any $2$-point Schwinger function with an inflexion point at $x=p^2>0$, Eq.\,(\ref{RPbroken}), must breach the axiom of reflection positivity.  This entails that the associated elementary excitation cannot appear in the Hilbert space of observables.  (Additional explanation can be found in Sec.\,2 of Ref.\,\cite{Roberts:2007ji}.)

\begin{figure}[t]
\begin{center}
\includegraphics[clip,width=0.33\textwidth,angle=-90]{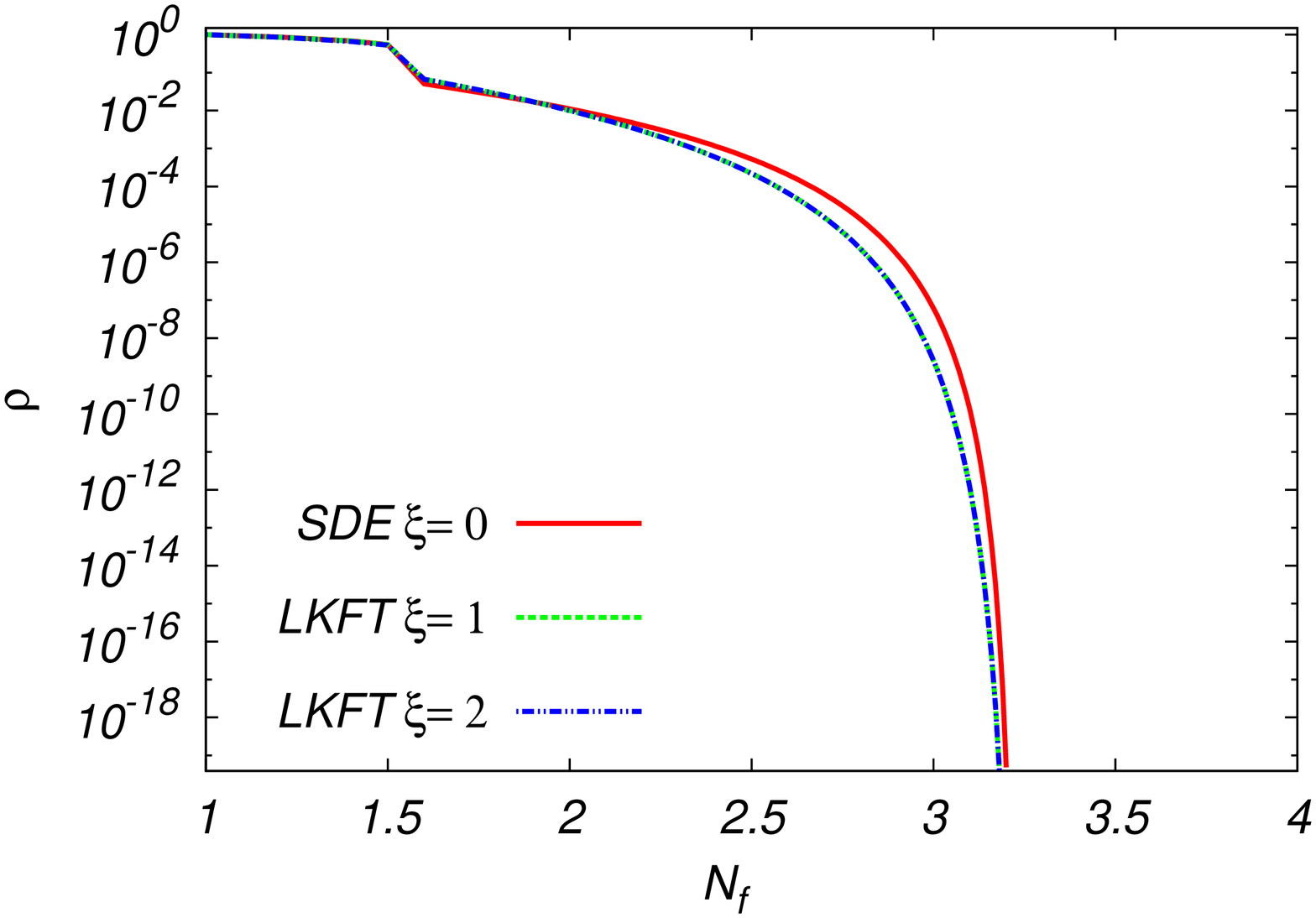}
\includegraphics[clip,width=0.33\textwidth,angle=-90]{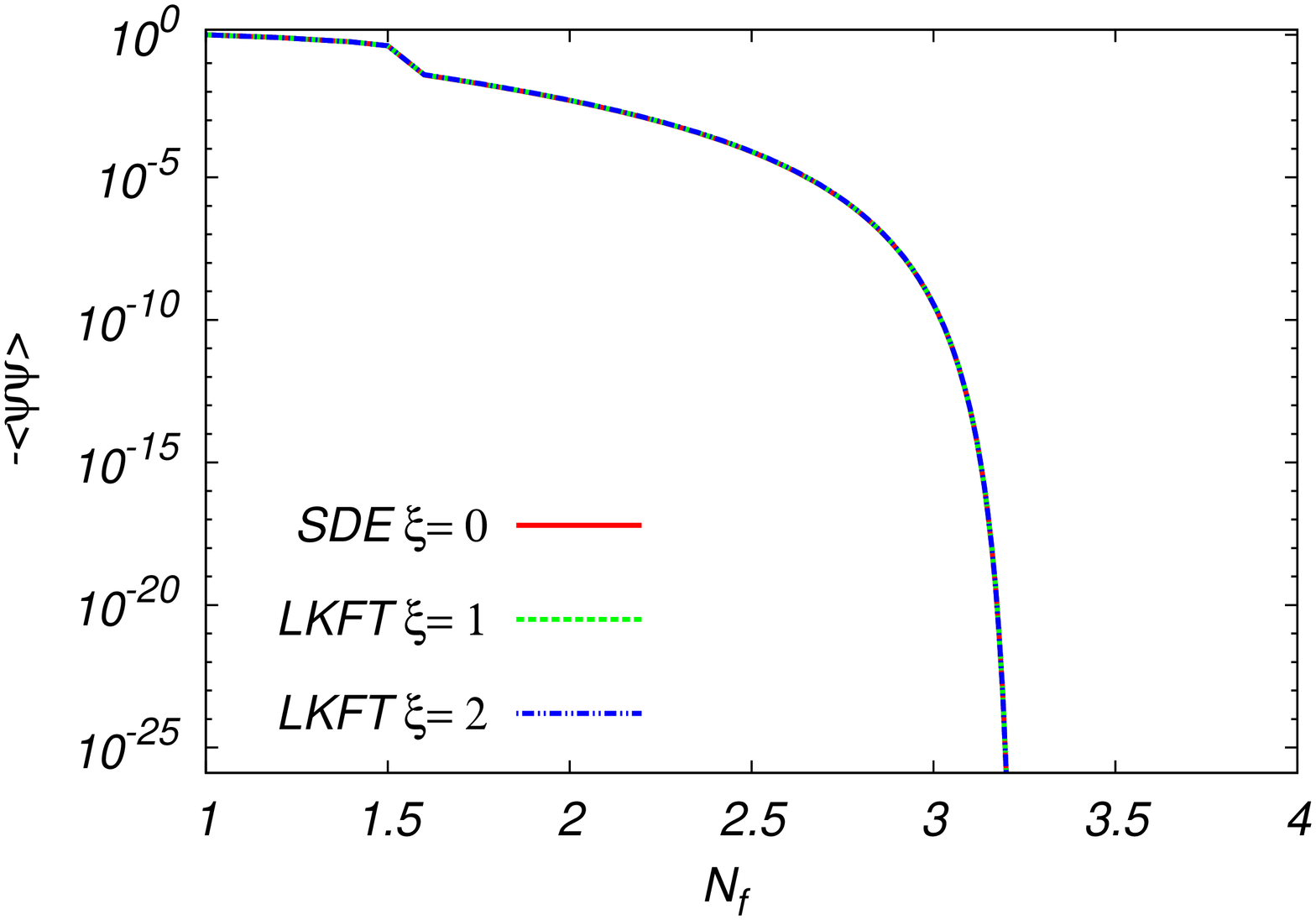}
\end{center}

\caption{\label{Fig1} \emph{Upper panel} -- $\rho$ in Eq.\,(\protect\ref{rhoDCSB}) for $\xi=0$; i.e., Landau gauge, and $\xi=1,2$ as obtained through the momentum space LKF transformation, Eq.\,(\protect\ref{gist}).  Plainly, the critical number of flavours for chiral symmetry restoration, $N_f^c$, is independent of $\xi$: in the model defined by Eqs.\,(\protect\ref{GammaW}) and (\protect\ref{CJBPolA}) we have $N_f^c=\frac{32}{\pi^2} = 3.24$ \cite{Bashir:2008fk}.    
\emph{Lower panel} -- Vacuum fermion condensate, Eq.\,(\protect\ref{qbarq}).  This quantity is manifestly independent of gauge parameter, as explicated in \protect\cite{Bashir:2005wt}.
(Remember, $e^2=1$ throughout.)}
\end{figure}

In order to illustrate general features of QED3, Ref.\,\cite{Bashir:2008fk} employed efficacious but simple models for the photon vacuum polarisation and fermion-photon vertex.  We follow suit, and use the so-called central Ball-Chiu vertex \cite{Ball:1980ay}
\begin{equation}
\label{GammaW}
\Gamma_\mu(p,q) = \frac{1}{2} \left[A(p)+A(q)\right] \gamma_{\mu} \;;
\end{equation}
and 
\begin{eqnarray}
\label{CJBPolA}
 \lefteqn{ \Pi(p,q) =   {\cal P}(p,q)\, \Pi(p-q)\,,} \\
\Pi(k) &= & N_f \left[\frac{1}{8} \frac{1}{\sqrt{k^2+ \eta a^2}} + \eta\,b \,{\rm e}^{- c k^2}\right],
\label{CJBPol}
\end{eqnarray}
with $\eta = \exp(-2(N_f-1)/\rho(N_f))$, and $a=0.20$, $b=0.088$ and $c=7.8$, which owe their nonzero values to DCSB and were calculated \cite{Burden:1991uh} for $N_f=1$.  (NB.\ For $a=0=b$, Eq.\,(\ref{CJBPol}) reduces to the leading order result in a $1/N_f$-expansion \cite{Appelquist:1988sr}.)

An extensive body of research, with the estimable goal of determining the optimal or even true form of the Landau-gauge vertex; e.g., Refs.\,\cite{Burden:1990mg,Burden:1993gy,Dong:1994jr,Bashir:1994az,Hawes:1996mw,%
Maris:1996zg,Bashir:2001vi,Kizilersu:2009kg},
has produced \emph{Ans\"atze} more sophisticated than Eq.\,(\ref{GammaW}).  Studies of the polarisation are also available \cite{Burden:1991uh,Maris:1996zg,Fischer:2004nq,Kizilersu:2009kg}.  Notwithstanding this, Eqs.\,(\ref{GammaW}) and (\ref{CJBPolA}) are sufficient herein.

\begin{figure}[t]
\vspace*{-2ex}

\centerline{\includegraphics[clip,width=0.4\textwidth,angle=-90]{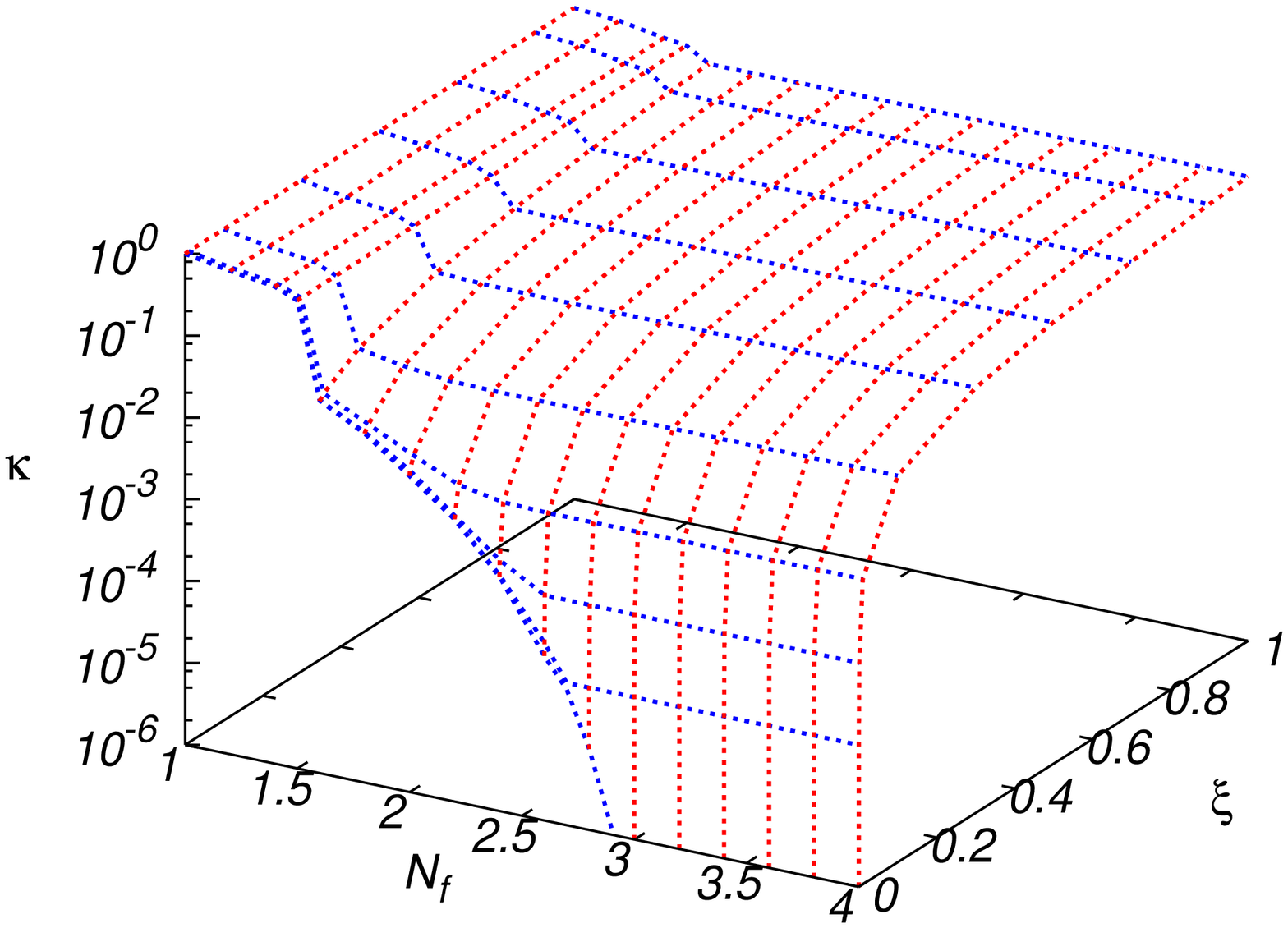}}
\centerline{\includegraphics[clip,width=0.35\textwidth,angle=-90]{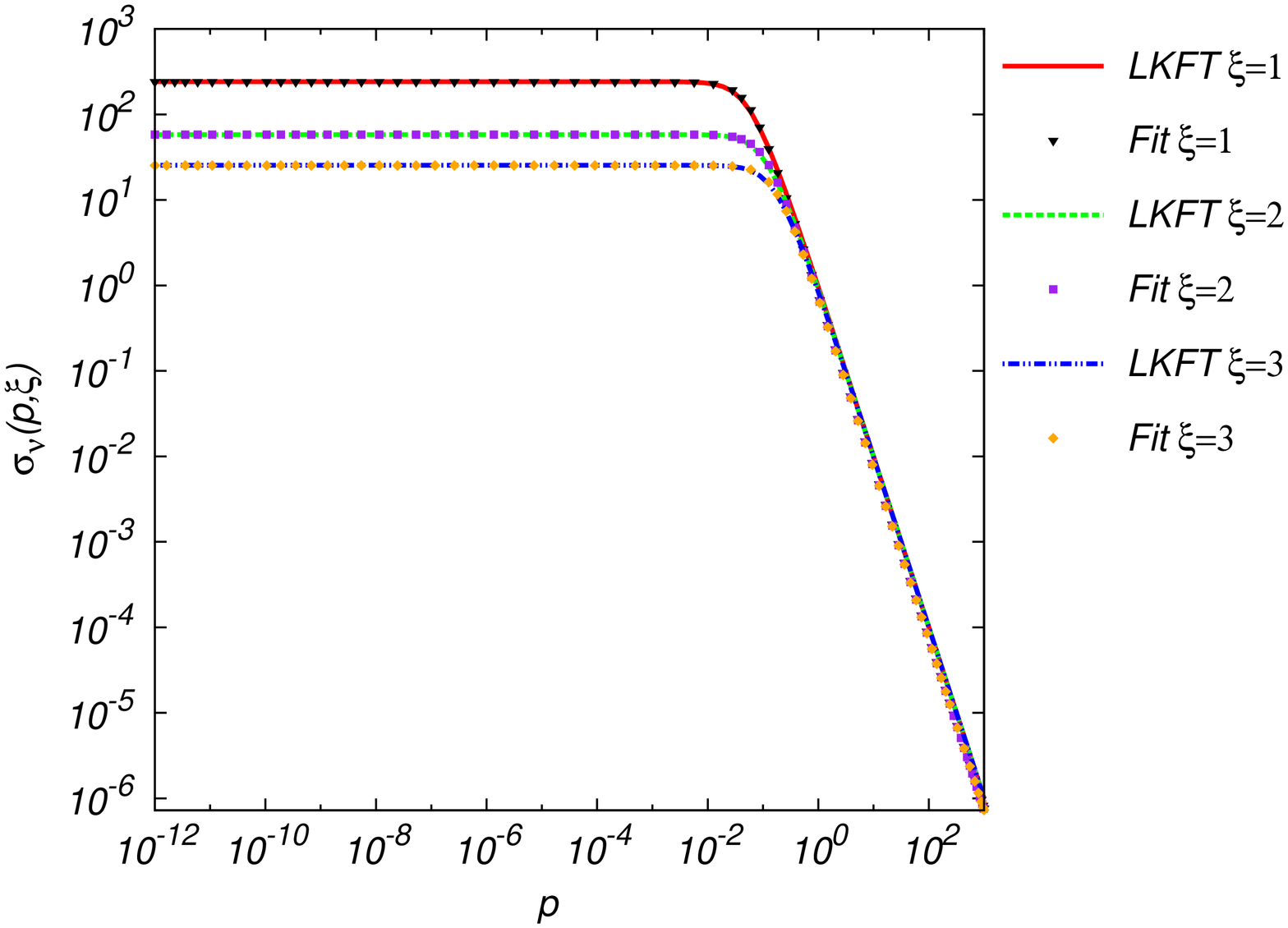}}

\caption{\label{Fig2} \emph{Upper panel} -- Confinement order parameter, $\kappa(N_f)$ in Eq.\,(\ref{Kconf}), and its dependence on the gauge parameter as determined by the LKF transformation, Eq.\,(\protect\ref{gistV}).  
\emph{Lower panel} -- The curves show $\sigma_{V}$ as a function of $p$, plotted for three nonzero values of the gauge parameter.  The symbols represent a fit to these functions with the form $z(\xi)/(\varsigma(\xi)^2+p^2)$, where $\varsigma(\xi)$ is precisely the LKF mass defined in Eq.\,(\protect\ref{LKFx}).}
\end{figure}

In Fig.\,\ref{Fig1} we plot the DCSB order parameter $\rho(N_f;\xi)$ as a function of $N_f$ in different $\xi>0$ gauges.\footnote{The curves in Fig.\,\protect\ref{Fig1} and the upper panel of Fig.\,\protect\ref{Fig2} exhibit a sudden drop at $N_f=1.55$.  As explained in Ref.\,\cite{Bashir:2008fk}, this artefact arises from the simple manner by which our model for the vacuum polarisation admits feedback from the fermion gap equation; i.e., through $\eta$.  It has no impact on our analysis.}  It is evident that this order parameter is only weakly sensitive to the gauge parameter and, furthermore, that the critical number of flavours for DCSB is independent of $\xi$.  In hindsight, this is obvious from Eq.\,(\ref{LKFx}) or (\ref{gist}).  To see this, consider the latter equation.  The kernel of the momentum-space LKF transformation is positive definite.  Hence a nonzero $\sigma_S(k)$ will remain nonzero after transformation.  On the other hand, if $\sigma_S(k;\xi=0)$ vanishes at and above some value of $N_f$, as it does when chiral symmetry is restored, then this feature is preserved by the LKF transformation.  

In Fig.\,\ref{Fig1} we also display the vacuum fermion condensate, Eq.\,(\protect\ref{qbarq}).  It is clearly independent of the gauge parameter, a result which is straightforward to establish when working with the LKF transformations in configuration space.  However, the manner by which this invariance is expressed in momentum space is interesting.  It is achieved through a redistribution of support in the fermion mass function: as $\xi$ increases from zero, strength at low momenta is shifted to larger momenta \cite{Bashir:2005wt}.

In the upper panel of Fig.\,\ref{Fig2} we depict the gauge-parameter-dependence of $\kappa(N_f)$ in Eq.\,(\ref{Kconf}).  At first glance this figure would appear to suggest that confinement is \emph{absent} in all but Landau gauge.  That conclusion, however, ignores two facts.  The first, remarked upon in the Introduction, is that any mass-scale generated dynamically in QED3 is at least one order of magnitude smaller than the natural scale, which is set by the dimensioned coupling $e^2=1$; and the second, noted in closing Sec.\,\ref{QED3recap}, is that the LKF transformations cannot generate a gauge-invariant pole mass.

The lower panel of Fig.\,\ref{Fig2} helps to resolve the conundrum.  Confinement is a long-range phenomenon.  It is expressed through $\kappa(N_f)$ in the location of the minimum of the first derivative of $\sigma_V(p^2)$: if there is more than one minimum, that with largest magnitude is chosen.  This minimum defines a confinement mass-scale, which by its nature must vanish as $N_f$ increases.  Hence, in the neighbourhood of $N_f^c$, for arbitrarily small values of the gauge parameter, the confinement mass-scale is overwhelmed by the LKF mass, $\varsigma(\xi)$ in Eq.\,(\ref{LKFx}).  Indeed, as the lower panel shows, the LKF mass very quickly comes to dominate completely the evolution of $\sigma_V$.  In these observations lies explanation of the behaviour apparent in the upper panel of Fig.\,\ref{Fig2}; namely, that with increasing $\xi$ the curves marking the $N_f$-dependence of $\kappa(N_f)$ depart from the Landau gauge result at smaller values of $N_f$.

For a given value of $\xi$ one could recover the signal of the confinement mass-scale by working with a higher-order derivative of $\sigma_V(p^2)$: higher-order derivatives amplify infrared effects and the order needed would depend on $\xi$.  Alternatively, one can switch to the confinement order parameter introduced in Ref.\,\cite{Hawes:1993ef}, following Ref.\,\cite{Hollenberg:1992nj}.  This order parameter is connected with the configuration space Schwinger function
\begin{equation}
\label{DeltaS}
\Delta_S(\tau) = \frac{1}{4} \int d^2 x \, {\rm tr}_{\rm D} S(x)\,.
\end{equation}
If this Schwinger function possesses a zero, then the axiom of reflection positivity is violated.  The location of the zero can therefore be connected with an order parameter for confinement: the feature is lost if the zero moves to $\tau=\infty$.  One can read from Eq.\,(\ref{LKFx}) that the LKF transformation does not introduce or eliminate zeros in $\Delta_S(\tau)$ because the LKF mass is purely real.

\begin{figure}[t]
\centerline{\includegraphics[clip,width=0.33\textwidth,angle=-90]{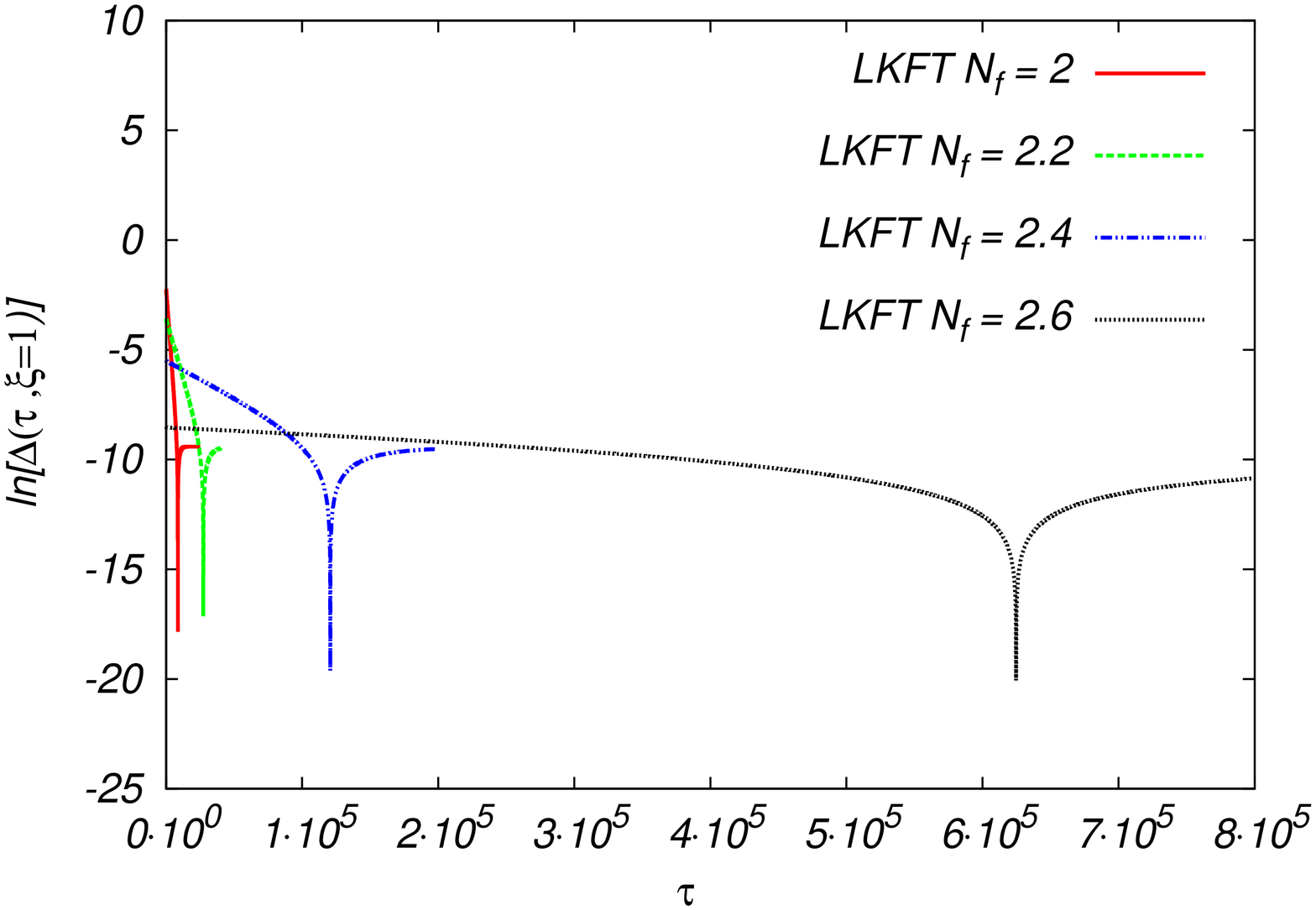}}
\centerline{\includegraphics[clip,width=0.33\textwidth,angle=-90]{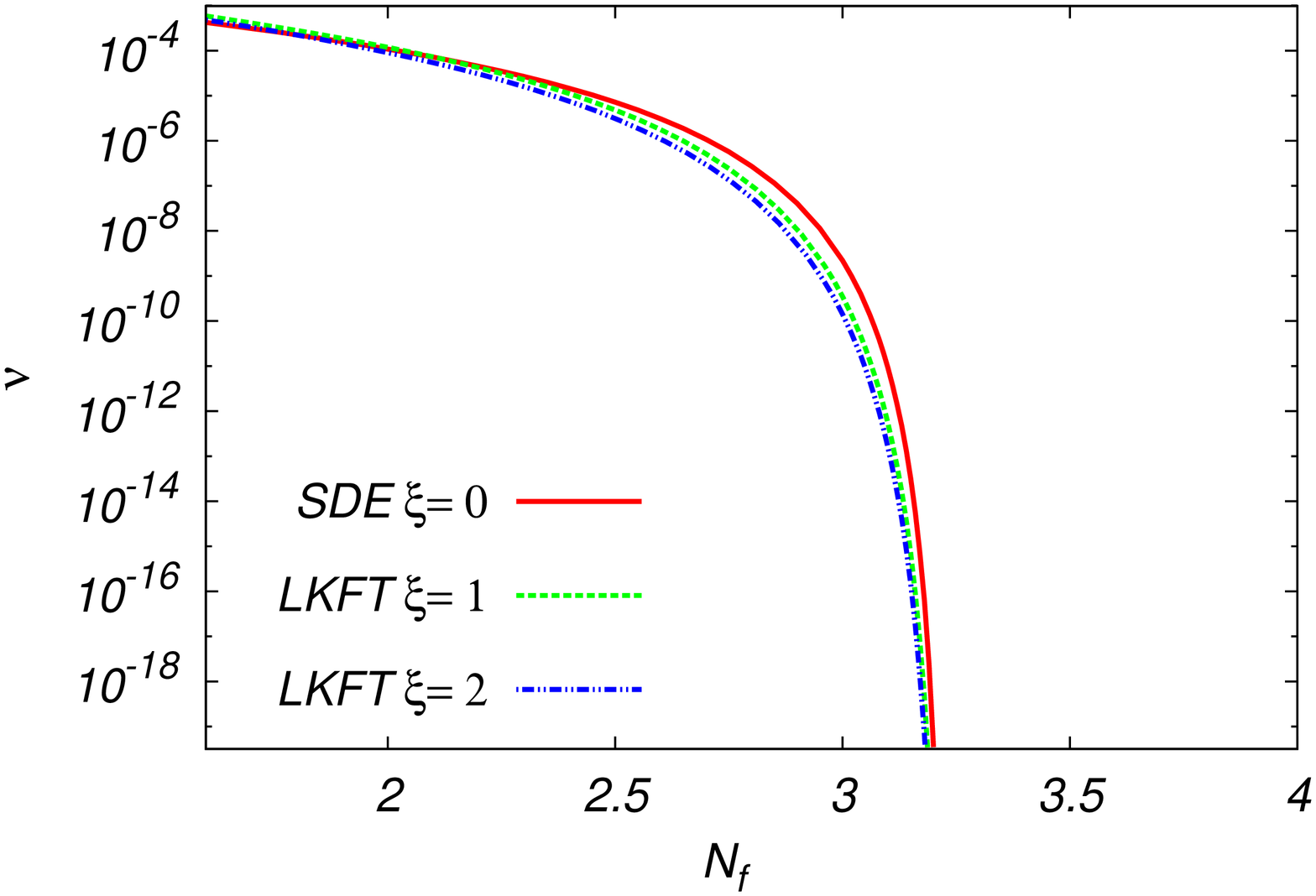}}

\caption{\label{Fig4} \emph{Upper panel} -- Dimensionless quantity ${\rm ln}|\Delta(\tau)|$, evaluated in Feynman gauge, $\xi=1$, as a function of Euclidean time for four values of $N_f<N_f^c$.  In order to preserve clarity, we cease plotting each curve at the first zero in $\Delta(\tau)$ but in each case $\Delta(\tau)$ exhibits periodic zeros, whose existence is tied to the presence in the momentum space propagator of a pole with a nonzero imaginary part, and its conjugate partner \protect\cite{Krein:1990sf,Hawes:1993ef}.
\emph{Lower panel} -- Confinement order parameter, $\nu(N_f;\xi$) in Eq.\,(\protect\ref{nuconf}), as a function of $N_f$ for three values of the gauge parameter.
}
\end{figure}

In the upper panel of Fig.\,\ref{Fig4} we plot $\ln |\Delta_S(\tau)|$, which exhibits periodic singularities $\forall\,N_f < N_f^c$: the singularities mark the zeros of $\Delta_S(\tau)$.  It is apparent that as $N_f \to N_f^c$ the location of the first zero moves to larger values of $\tau$.  As an order parameter for confinement we therefore employ \cite{Bender:1996bm}
\begin{equation}
\label{nuconf}
\nu(N_f) := \frac{1}{\tau_1(N_f)},
\end{equation}
where $\tau_1(N_f)$ is the location of the first zero.  This order parameter vanishes when confinement is lost.  It is notable that the magnitude of the curves, which is connected with $\rho(N_f)$ in Eq.\,(\ref{rhoDCSB}), also diminishes rapidly as $N_f \to N_f^c$.  This highlights the intimate connection between chiral symmetry restoration and deconfinement.

We depict the confinement order parameter in the lower panel of Fig.\,\ref{Fig4}.  It is clear that there is a critical number of flavours, $N_{f\nu}^c$, above which confinement is lost.  This critical value is independent of the gauge parameter, a feature that was anticipated and explained after Eq.\,(\ref{DeltaS}).  Moreover, $N_{f\nu}^c=N_f^c$; i.e., chiral symmetry restoration and deconfinement are simultaneous in our illustrative model.  The fundamental causal connection is a dramatic change in the analytic properties of the propagator that accompanies the disappearance of a nonzero fermion scalar self-energy.  As we explained above, the LKF transformation cannot materially affect this.

\section{Conclusion}
\label{sec:close}
A transition between elements in the class of covariant gauges is effected by a Landau-Khalatnikov-Fradkin (LKF) transformation.  The action of this transformation on the fermion propagator in noncompact QED3 can be written in a closed form in both configuration and momentum space.  Therefore it is always possible to anticipate a range of effects that the gauge transformation can have on the propagator's analytic properties.  We used these facts to argue that if a critical number of flavours for chiral symmetry restoration and deconfinement exists in QED3, then its value is independent of the gauge parameter.  We explicated these arguments with the aid of simple models for the fermion-photon vertex and photon vacuum polarisation.  In doing so we highlighted pitfalls, a failure of which to avoid will lead to erroneous conclusions.  

In focusing on the LKF transformations we were led anew to the view that Landau gauge occupies a special place.  In addition to other important properties, such as being a fixed point of the renormalisation group and the gauge in which any sensitivity to model-dependent differences between \emph{Ans\"atze} for the fermion-photon vertex are least noticeable, one can add that it is the sole covariant gauge in which the infrared behaviour of the fermion propagator is not modified by a non-dynamical gauge-dependent exponential factor whose presence can obscure truly observable features of the theory. Our model was helpful in elucidating this. 

With our \emph{Ans\"atze} for the fermion-photon vertex and photon vacuum polarisation, QED3 exhibits chiral symmetry restoring and deconfinement transitions when the number of flavours exceeds a common critical value.  We expect this simultaneity to persist in QED3 proper, so long as chiral symmetry is dynamically broken in the quenched massless theory.  Our presumption is based upon Landau gauge Dyson-Schwinger equation (DSE) studies of QCD in rainbow-ladder truncation, which find these transitions to be coincident at $T\neq 0$ \protect\cite{Bender:1996bm} and $\mu \neq 0$ \protect\cite{Bender:1997jf,Chen:2008zr}; and Ref.\,\cite{Blaschke:1997bj}, which describes a model that exhibits a line of simultaneous transitions in the physical quadrant of the $(T,\mu)$-plane.  Indeed, chiral symmetry restoration and deconfinement are coincident in all careful self-consistent studies of concrete models of continuum chiral-limit QCD that exhibit both phenomena. This is also the case in numerical simulations of lattice-regularised QCD \protect\cite{Sinclair:2008du}.

Our study supports a view that results derived from Landau gauge analyses will persist in any covariant gauge.  Nonetheless, this does not obviate the need for discovering a quantitatively reliable Landau-gauge DSE truncation \cite{Binosi:2007pi,Chang:2009zb,Kizilersu:2009kg}.  

\begin{acknowledgments}
\hspace*{-\parindent}\parbox{24em}{%
CDR is grateful to the staff and students of the \emph{Instituto de F\'{\i}sica y Matem\'aticas, Universidad Michoacana de San Nicol\'as de Hidalgo}, for organising and hosting the ``2nd Morelia Workshop on Nonperturbative Aspects of Field Theories,'' at which this study was completed.
This work was supported by:
CIC and CONACyT grants, under project nos.\ 4.10, 4.22, 82230 and 46614-I; COECyT grants;
\vspace*{\fill}
and the Department of Energy, Office of Nuclear Physics, contract no.~DE-AC02-06CH11357.
}
\end{acknowledgments}
\vspace*{\fill}

\bibliography{LKFQED3}

\end{document}